\documentclass[twocolumn,aps,tightenlinefs,superscriptaddress,preprintnumbers]{revtex4}

\usepackage{graphicx,color,dcolumn,booktabs,bm}
\usepackage{longtable,lscape}
\usepackage{amsmath}
\usepackage{indentfirst}
\usepackage{epsfig}
\usepackage{feynmf}   
\usepackage{epstopdf}   
\usepackage{slashed}  
\usepackage{cases}
\usepackage{color}
\usepackage{multirow}
\usepackage{ulem}
\usepackage{verbatim}
\usepackage[colorlinks,linkcolor=red,anchorcolor=green,citecolor=blue]{hyperref}
\usepackage{mathrsfs}
\usepackage{rotating}
\usepackage{threeparttable}
\usepackage{lineno}
\usepackage{subfigure}
\usepackage{gensymb}
\usepackage{microtype}

\graphicspath{{ps}}

\newcommand{\etacp}{\eta_c(2S)}

\newcommand{\pptripi}{p\bar{p}\pi^+\pi^-\pi^0}
\newcommand{\BF}{\mathcal{B}}

\begin{document}

\title{ \quad\\[0.1cm] \boldmath  Evidence for $\eta_{c}(2S)\to p\bar{p}\pi^{+}\pi^{-}\pi^{0}$ and observation of $\chi_{cJ} \to p\bar{p}\pi^{+}\pi^{-}\pi^{0}$}

\newcommand{\BESIIIorcid}[1]{\href{https://orcid.org/#1}{\hspace*{0.1em}\raisebox{-0.45ex}{\includegraphics[width=1em]{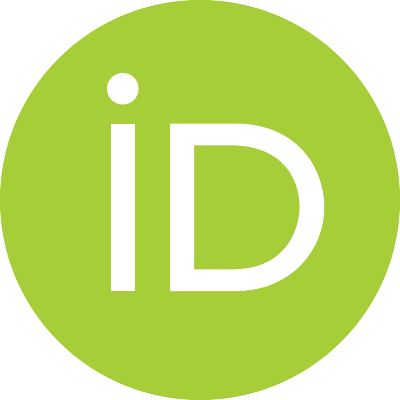}}}}

\author{
M.~Ablikim$^{1}$\BESIIIorcid{0000-0002-3935-619X},
M.~N.~Achasov$^{4,c}$\BESIIIorcid{0000-0002-9400-8622},
P.~Adlarson$^{83}$\BESIIIorcid{0000-0001-6280-3851},
X.~C.~Ai$^{89}$\BESIIIorcid{0000-0003-3856-2415},
C.~S.~Akondi$^{31A,31B}$\BESIIIorcid{0000-0001-6303-5217},
R.~Aliberti$^{39}$\BESIIIorcid{0000-0003-3500-4012},
A.~Amoroso$^{82A,82C}$\BESIIIorcid{0000-0002-3095-8610},
Q.~An$^{78,65,\dagger}$,
Y.~H.~An$^{89}$\BESIIIorcid{0009-0008-3419-0849},
M.~S.~Anderson$^{39}$\BESIIIorcid{0009-0008-1550-2632}
% ...
\\
\vspace{0.2cm}
(BESIII Collaboration)\\
\vspace{0.2cm} {\it
$^{1}$ Institute of High Energy Physics, Beijing 100049, People's Republic of China\\
% ...
$^{4}$ Budker Institute of Nuclear Physics SB RAS (BINP), Novosibirsk 630090, Russia\\
% ...
$^{31}$ INFN Sezione di Ferrara, (A)INFN Sezione di Ferrara, I-44122, Ferrara, Italy; (B)University of Ferrara, I-44122, Ferrara, Italy\\
% ...
$^{39}$ Johannes Gutenberg University of Mainz, Johann-Joachim-Becher-Weg 45, D-55099 Mainz, Germany\\
% ...
$^{65}$ State Key Laboratory of Particle Detection and Electronics, Beijing 100049, Hefei 230026, People's Republic of China\\
% ...
$^{78}$ University of Science and Technology of China, Hefei 230026, People's Republic of China\\
% ...
$^{82}$ University of Turin and INFN, (A)University of Turin, I-10125, Turin, Italy; (B)University of Eastern Piedmont, I-15121, Alessandria, Italy; (C)INFN, I-10125, Turin, Italy\\
$^{83}$ Uppsala University, Box 516, SE-75120 Uppsala, Sweden\\
% ...
$^{89}$ Zhengzhou University, Zhengzhou 450001, People's Republic of China\\
\vspace{0.2cm}
$^{\dagger}$ Deceased\\
% ...
$^{c}$ Also at the Novosibirsk State University, Novosibirsk, 630090, Russia\\
% ...
}
%% ends here %%	
}

\begin{abstract}
 Using $(2.712\pm0.014)\times 10^9$ $\psi(3686)$ events collected by the BESIII detector at the BEPCII collider, the $\psi(3686) \to \gamma p\bar{p}\pi^+\pi^-\pi^0$ process is investigated. Evidence for the decay of $\eta_{c}(2S)\to p\bar{p}\pi^{+}\pi^{-}\pi^{0}$ is found with a signal significance of 3.3$\sigma$. The product of branching fractions of $\mathcal{B}[\psi(3686)\to \gamma \eta_{c}(2S)]\times\mathcal{B}[\eta_{c}(2S)\to  p\bar{p}\pi^{+}\pi^{-}\pi^{0}]$ is determined to be $(3.4\pm0.5\pm0.8) \times 10^{-6}$, where the first uncertainty is statistical and the second systematic. The hadronic decays of $\chi_{cJ} \to p\bar{p}\pi^+\pi^-\pi^0$$~(J=0,1,2)$ are observed, and their branching fractions are measured to be $\mathcal{B}(\chi_{c0}\to p\bar{p}\pi^{+}\pi^{-}\pi^{0})=(4.79\pm 0.01\pm0.40) \times 10^{-3}$, $\mathcal{B}(\chi_{c1}\to p\bar{p}\pi^{+}\pi^{-}\pi^{0})=(2.13\pm 0.01\pm0.17) \times 10^{-3}$, and $\mathcal{B}(\chi_{c2}\to p\bar{p}\pi^{+}\pi^{-}\pi^{0})=(3.72\pm 0.01\pm0.29) \times 10^{-3}$, respectively. 
 Furthermore, the branching fractions for the intermediate processes $\chi_{cJ}\to p\bar{p}\omega$ are updated with significantly improved precision: $\mathcal{B}(\chi_{c0}\to p\bar{p}\omega)=(5.76\pm0.01\pm0.42)\times10^{-4}$, $\mathcal{B}(\chi_{c1}\to p\bar{p}\omega)=(1.85\pm0.01\pm0.13)\times10^{-4}$, and $\mathcal{B}(\chi_{c2}\to p\bar{p}\omega)=(4.51\pm0.01\pm0.33)\times10^{-4}$, respectively.
\end{abstract} 

\maketitle

\tighten

\section{Introduction}
 The properties of charmonia below the open-charm threshold, particularly the spin-singlets including the $S$-wave ground state $\eta_c(1S)$ and its first radial excitation $\eta_c(2S)$, as well as the $P$-wave spin-triplets $\chi_{cJ}(J=0,1,2)$,  are not fully understood~\cite{pdg,intro1,intro2,intro3,intro4,intro5}. (Throughout this paper, $\chi_{cJ}$ refers to $\chi_{c0}$, $\chi_{c1}$, and $\chi_{c2}$.)

 The study of $\eta_c(2S)$ decays plays a crucial role in testing quantum chromodynamics (QCD) in the charmonium sector. Compared with its well-measured spin-triplet counterpart $\psi(3686)$, $\eta_c(2S)$ has poorly determined properties since its discovery~\cite{Hist1}. For example, the measured partial widths account for less than 10\% of the total width~\cite{pdg}. Furthermore, the study of $\eta_c(2S)$ decays provides valuable information for charmonium puzzles. Perturbative QCD predicts that the $J/\psi$ and its first radial excitation $\psi(3686)$ decay into any exclusive light-hadron final state with a width proportional to the square of the wave function at the origin~\cite{bfr1,555}, which is known as the ``12\% rule''. As spin partners of $J/\psi$ and $\psi(3686)$, the $\eta_c(1S)$ and $\eta_c(2S)$ can decay into light hadrons in a similar way. Theoretical models predict a ratio of 12\%~\cite{bfr3} or 100\%~\cite{bfr4}. However, most of the experimental measurements agree with neither prediction~\cite{bfr5}. Any search for new decay modes, such as $\eta_c(2S) \to p\bar{p}\pi^+\pi^-\pi^0$, provides direct insight into its decay mechanisms and helps to test the predictions from QCD. 

 The cross section of the charmonium production process $p\bar{p} \to \Psi M$ is of great interest for planned $p\bar{p}$ experiments, where $\Psi$ represents any $c\bar{c}$ bound state and $M$ represents a light meson. The partial width of $\Psi\to p\bar{p}M$ serves as a key input for calculating the charmonium production cross section in $p\bar{p}\rightarrow\Psi M$~\cite{Barnes1,Barnes2,Barnes3,Barnes4}. As a specific example, the decay $\chi_{cJ} \to p\bar{p}\omega$ has been observed by the CLEO Collaboration~\cite{cleoppomg}. However, the non-resonant process $\chi_{cJ} \to p\bar{p}\pi^+\pi^-\pi^0$ remains unmeasured. Precise measurements of $\chi_{cJ} \to p\bar{p}\omega$ and $\chi_{cJ} \to p\bar{p}\pi^+\pi^-\pi^0$ provide information that helps to understand the properties of $\chi_{cJ}$.

 In this paper, we report a search for the decay $\eta_{c}(2S) \to p\bar{p}\pi^{+}\pi^{-}\pi^{0}$ via the radiative transition $\psi(3686)\to \gamma\eta_c(2S)$, based on a data sample of $(2.712\pm0.014)\times 10^9$ $\psi(3686)$ events collected with the BESIII detector~\cite{psipNum}. The corresponding decay modes $\chi_{cJ} \to p\bar{p}\pi^+\pi^-\pi^{0}$ are also studied. Furthermore, improved measurements of branching fractions for $\chi_{cJ}\to p\bar{p}\omega$ with $\omega\to\pi^{+}\pi^{-}\pi^{0}$ are presented.

\section{BESIII detector and Monte Carlo simulation}
The BESIII detector~\cite{bes3detec} records symmetric $e^+ e^-$ collisions provided by the BEPCII storage ring~\cite{bepc} in the center-of-mass energy ($\sqrt{s}$) range from 1.84 to 4.95~GeV, with a peak luminosity of $1.1\times 10^{33}$~cm$^{-2}$s$^{-1}$ achieved at $\sqrt{s}=3.773$~GeV. BESIII has collected large data samples in this energy region~\cite{dataset1,dataset2,dataset3}. The cylindrical core of the BESIII detector covers 93\% of the full solid angle and consists of a helium-based multilayer drift chamber (MDC), a time-of-flight system (TOF), and a CsI(Tl) electromagnetic calorimeter (EMC), all enclosed in a superconducting solenoidal magnet providing a 1.0~T magnetic field. The solenoid is supported by an octagonal flux-return yoke with resistive plate counter muon identification modules interleaved with steel. The charged-particle momentum resolution at 1~GeV/$c$ is 0.5\%, and the d$E/$d$x$ resolution is 6\% for electrons from Bhabha scattering. The EMC measures photon energies with a resolution of 2.5\% (5\%) at 1~GeV in the barrel (end-cap) region. The time resolution in the plastic scintillator TOF barrel region is 68~ps, while that in the end cap region is 110~ps. The end-cap TOF system was upgraded in 2015 using multigap resistive plate chamber technology, providing a time resolution of 60~ps, which benefits 83\% of the data used in this analysis~\cite{besdet1,besdet2,besdet3}.

Monte Carlo (MC) simulated data samples are used to determine detection efficiencies and to estimate backgrounds. They are produced with a {\sc geant4}-based software package~\cite{geant4}, which includes the geometric description of the BESIII detector and the detector response. The simulation models the beam-energy spread and initial-state radiation (ISR) in the $e^+e^-$ annihilations with the generator {\sc kkmc}~\cite{kkmc1,kkmc2}. The inclusive MC sample includes the production of the $\psi(3686)$ resonance, the ISR production of the $J/\psi$, and the continuum processes incorporated in {\sc kkmc}~\cite{kkmc1,kkmc2}. All particle decays are modeled with {\sc evtgen}~\cite{evtgen1,evtgen2} using branching fractions (BFs) either taken from the Particle Data Group (PDG)~\cite{pdg}, when available, or otherwise estimated with {\sc lundcharm}~\cite{lundcharm1,lundcharm2}. Final-state radiation (FSR) from charged final-state particles is incorporated using the {\sc photos} package~\cite{photos}. An inclusive MC sample of $2.747\times10^9$ $\psi(3686)$ events is used and normalized to the same integrated luminosity as the dataset.

The exclusive signal decays of $\psi(3686) \to \gamma \eta_c(2S)$ and $\psi(3686) \to \gamma \chi_{cJ}$ are generated based on their helicity decay amplitudes~\cite{chicj}, which follow an angular distribution of $(1+\lambda\cos^2\theta^{*}_{\gamma})$, where $\theta^{*}_{\gamma}$ denotes the polar angle of the radiated photon in the $\psi(3686)$ rest frame. The parameter $\lambda$ takes the value of $1$ for $\eta_c(2S)$, and the values of $1$, $-1/3$, and $1/13$ for $\chi_{cJ}$ with $J=0$, $1$, and $2$, respectively~\cite{chicj}. The subsequent decays $\eta_c(2S)/\chi_{cJ} \to \pptripi$ and $p\bar{p}\omega$ are generated uniformly in phase space and corrected using the momentum distributions of final states and mass spectra related to intermediate states for the $\chi_{cJ}$ decays. A Dalitz model is used for the decay $\omega \to \pi^+\pi^-\pi^0$~\cite{omgdalitz}.  
An exclusive background MC sample for the decay $\psi(3686) \to \pptripi(\gamma_\text{FSR})$ is generated with a phase-space model to estimate the background contribution. Separate exclusive MC samples, each containing $3\times10^6$ $\psi(3686)$ events, are used.
  
In addition, the data sample collected at $\sqrt{s}=3.650~\text{GeV}$~\cite{psipNum}, corresponding to an integrated luminosity of 445.5 $\text{pb}^{-1}$, is used to estimate the continuum background contribution.     

%%%%%%%%%%%%%%%%%%%%%%%%%%%%%%%%%%%%%%%%%%%%%%%%%%%%%%%%%%%%%%%%%%%%%
\section{Event Selection}
 A search for $\etacp$ is conducted through the exclusive decay $\psi(3686) \to \gamma\etacp$ with $\etacp \to \pptripi$ and $\pi^0 \to \gamma\gamma$, comprising three photons and four charged particles in the final states. 

 Tracks detected in the MDC must be within a polar angle ($\theta$) range of $|\cos\theta|<0.93$, where $\theta$ is defined with respect to the symmetry axis of the MDC ($z$-axis). For tracks, the distance of closest approach to the interaction point (IP) must be less than 10\,cm along the $z$-axis and less than 1\,cm in the transverse plane. Only events with exactly four tracks with net charge zero are selected. Particle identification (PID) combines information from energy deposited in the MDC~(d$E$/d$x$) and the flight time in the TOF to compute likelihoods $\mathcal{L}(h)$ for each hadron hypothesis $h$ ($h=\pi,~K,~p$). A track is identified as a pion if it satisfies $\mathcal{L}(\pi)>\mathcal{L}(K)$, $\mathcal{L}(\pi)>\mathcal{L}(p)$, and $\mathcal{L}(\pi)>0.001$. Proton candidates are identified by imposing $\mathcal{L}(p)>\mathcal{L}(\pi)$, $\mathcal{L}(p)>\mathcal{L}(K)$, and $\mathcal{L}(p)>0.001$.

 Photon candidates are identified using showers in the EMC. Showers are selected by requiring a minimum deposited energy of 25~MeV in both the barrel region ($|\cos \theta|< 0.80$) and end-cap region ($0.86 <|\cos \theta|< 0.92$). To suppress electronic noise and showers unrelated to the event, the difference between the EMC time and the event start time is required to be within [0, 700]\,ns. The number of photon candidates is required to be greater than three.

The $\pi^0$ candidates are reconstructed from photon pairs. A one-constraint kinematic fit is applied to each $\gamma\gamma$ combination, where the invariant mass of the $\gamma\gamma$ pair is constrained to the known $\pi^0$ mass. The fit must converge with a $\chi^2$ value less than 200. 
  
The four charged tracks are constrained to originate from a common vertex. To suppress background events from long-lived particles, the vertex fit is required to converge with a $\chi^2$ value less than 100. A four-constraint (4C) kinematic fit~\cite{4cfit} is performed by constraining the total four-momentum of the final-state particles to the initial $\psi(3686)$ four-momentum. For events with multiple radiative photon and $\pi^0$ candidates (where a radiative photon is defined as one not originating from a $\pi^0$ decay), the $\gamma p\bar{p}\pi^+\pi^-\pi^0$ combination with the minimum smallest chi-squared value from the 4C kinematic fit $(\chi^2_\text{4C})$ is retained. The requirement $\chi^2_\text{4C}<40$ is applied, a value determined through optimization by maximizing the figure of merit $S/\sqrt{S+B}$~\cite{fom}, where $S$ and $B$ are the expected numbers of signal and background events, estimated from the signal MC sample with the measured branching fraction and normalized inclusive MC samples, respectively. Furthermore, the invariant mass of the photon pair is required to lie within the optimized mass window of (0.115, 0.150)~GeV/$c^2$.

Two additional 4C kinematic fits are performed under the $\psi(3686)\to\gamma\gamma p\bar{p}\pi^+\pi^-$ and $\psi(3686)\to\gamma\gamma\gamma\gamma p\bar{p}\pi^+\pi^-$ hypotheses. The candidate is kept only if $\chi^2_\text{4C}<\chi^2_\text{4C}(2\gamma)$ and $\chi^2_\text{4C}<\chi^2_\text{4C}(4\gamma)$ to suppress background events with unexpected photon multiplicities. Here, $\chi^2_\text{4C}(2\gamma)$ and $\chi^2_\text{4C}(4\gamma)$ represent the $\chi^2$ value for $\psi(3686)\to\gamma\gamma p\bar{p}\pi^+\pi^-$ and $\psi(3686)\to\gamma\gamma\gamma\gamma p\bar{p}\pi^+\pi^-$ hypotheses, respectively.

To suppress cross-feed between the radiative photon and daughter photons from $\pi^0$ decays, combinations of the radiative photon and $\pi^0$ daughter photons must satisfy $\left|M(\gamma_{\pi^0,1}\gamma_\text{rad})-m(\pi^0)\right|$~$>15$~MeV$/c^2$  and $\left|M(\gamma_{\pi^0,2}\gamma_\text{rad})-m(\pi^0)\right|$~$>12$~MeV$/c^2$, where $M(\gamma_{\pi^0,1(2)}\gamma_\text{rad})$ is the invariant mass of the $\gamma_{\pi^0,1(2)}\gamma_\text{rad}$ combination and $m(\pi^0)$ is the known $\pi^0$ mass~\cite{pdg}. The mass windows are determined from signal MC simulation.

%%%%%%%%%%%%%%%%%%%%%%%%%%%%%%%%%%%%%%%%%%%%%%%%%%%%%%%%%%%%%%%%%%%%%
\section{Background estimation}
\label{bkgest}

A study with the TopoAna tool~\cite{topoana} on the inclusive MC sample reveals that the dominant background events are from the decays $\psi(3686) \to \pi^+\pi^-J/\psi$ and $\psi(3686) \to \pi^{0}\pi^{0} J/\psi$. Background events from $\psi(3686) \to \pi^+\pi^-J/\psi$ are excluded by requiring the recoil mass ($RM$) of the $\pi^+\pi^-$ pair to satisfy $\left|RM(\pi^{+}\pi^{-})-m(J/\psi)\right|>6$~MeV/$c^2$, where $m(J/\psi)$ is the known $J/\psi$ mass~\cite{pdg}. Background events from $\psi(3686) \to \pi^{0}\pi^{0} J/\psi, J/\psi \to p\bar{p}\pi^{+}\pi^{-}$ are discarded by requiring the invariant mass of the $p\bar{p}\pi^{+}\pi^{-}$ system to satisfy $|M(p\bar{p}\pi^{+}\pi^{-})-m(J/\psi)|>23$~MeV/$c^2$. 
Background events from the decay $\psi(3686)\to\Sigma^0\bar{\Lambda}\pi^0$, as well as other decays proceeding through hyperon intermediate states such as $\Lambda$ and $\Sigma$, are rejected by requiring $M(p\pi^-) \notin (1.103,~1.121)$~GeV/$c^2$, $M(p\pi^0) \notin (1.172,~1.200)$~GeV/$c^2$, and $M(\gamma\Lambda) \notin (1.178,~1.202)$~GeV/$c^2$. Here, $M(p\pi^-)$, $M(p\pi^0)$, and $M(\gamma\Lambda)$ are the invariant masses of the $p\pi^-$, $p\pi^0$, and $\gamma\Lambda$ combinations, respectively.

The other dominant background stems from direct decay of $\psi(3686) \to \pptripi(\gamma_{\text{FSR}})$, where either a fake photon or an FSR photon is incorporated into the reconstructed final state. The inclusion of a fake photon in the 4C kinematic fit causes the $\pptripi$ invariant mass to peak in the vicinity of $\etacp$ rather than at the $\psi(3686)$ mass. To mitigate this effect, a three-constraint (3C) kinematic fit~\cite{3cfit} is applied, in which the energy of the radiative photon is left unconstrained. This procedure shifts the background peak back to the $\psi(3686)$ mass region, while leaving the signal peak unchanged. Inclusion of the FSR photon introduces a long tail in the $\pptripi$ invariant-mass [$M(\pptripi)$] distribution, contaminating the $\etacp$ signal. The exclusive MC sample of $\psi(3686) \to \pptripi(\gamma_\text{FSR})$ is used to model this background. The FSR correction factor $f_\text{FSR} = 1.69 \pm 0.05$, with statistical uncertainty dominant, is applied to account for the data-MC difference in the probability of FSR photon emission, which is determined from a control sample of $\psi(3686)\to\gamma\chi_{c0},~\chi_{c0} \to p\bar{p}\pi^+\pi^-(\gamma_\text{FSR})$ using a method similar to that reported in Ref.~\cite{3cfit}. The impact of a fake photon on the estimation of the FSR correction factor is negligible.

Background events involving misreconstructed radiative photons contribute a broad peak that contaminates the $\etacp$ signal. Their contribution is investigated using signal MC samples of $\psi(3686)\to\gamma\chi_{cJ},~\chi_{cJ}\to\pptripi$. Simulated events are classified as misreconstructed background events if they satisfy the requirements $\theta_{\rm mis\gamma}>5^{\degree}$ for $\chi_{c0,2}$ and $\theta_{\rm mis\gamma}>6^{\degree}$ for $\chi_{c1}$, where $\theta_{\rm mis\gamma}$ is defined as the opening angle between the reconstructed and generated radiative photons. 

Continuum background events (including initial state radiation) are estimated using the dataset recorded at $\sqrt{s}=3.650$~GeV. The momenta and energies of final-state particles are scaled to account for the difference in $\sqrt{s}$. The mass spectrum is normalized by a factor $f_\text{norm}=(\mathcal{L}_{\sqrt{s}=3.686}/\mathcal{L}_{\sqrt{s}=3.650})\cdot(\frac{1}{3.686^2}/\frac{1}{3.650^2})$ based on the differences in integrated luminosity ($\mathcal{L}$) and the energy-dependent cross section (assumed to be proportional to $1/s$). 
%%%%%%%%%%%%%%%%%%%%%%%%%%%%%%%%%%%%%%%%%%%%%%%%%%%%%%%%%%%%%%%%%%%%%
\section{branching fraction measurement}
\subsection{\boldmath \texorpdfstring{$\eta_c(2S)\to \pptripi$ and $\chi_{cJ}\to \pptripi$}{Lg}}
 The signal yields of $\eta_c(2S)\to \pptripi$ and $\chi_{cJ}\to \pptripi$ are extracted by performing an unbinned maximum likelihood fit to the $\pptripi$ invariant mass spectrum in data after imposing event selections and applying the 3C kinematic fit.  

 The signal lineshape is modeled by 
 \begin{linenomath*}
 \begin{equation} 
     \label{sigsp}
     [E_{\gamma}^{3}\cdot BW(m) \cdot f_{d}(E_{\gamma})\cdot \epsilon(m) ]\otimes G(\delta_{m},\delta_{\sigma}), 
 \end{equation}
 \end{linenomath*}
where $m$ denotes the invariant mass of $\pptripi$. The factor $E^3_{\gamma}$ accounts for the $M1$ transition contribution to the decay width~\cite{cleowork}, with $E_{\gamma}= [m_{\psi(3686)}^2-m^2]/[2\cdot m_{\psi(3686)}]$ being the energy of the transition photon in the $\psi(3686)$ rest frame. A Breit-Wigner (BW) function $BW(m)$ is used, with the masses and widths of $\eta_{c}(2S)$ and $\chi_{cJ}$ fixed to the world-averaged values~\cite{pdg}. To suppress the diverging tail arising from the $E_{\gamma}^3$ term, a damping factor $f_d(E_{\gamma})=E_0^2/[E_{\gamma}E_0 + (E_{\gamma}-E_0)^2]$ is introduced~\cite{kedr}, where $E_0$ is the most probable energy of the transition photon. The mass-dependent detection efficiency $\epsilon(m)$ is parameterized with an Argus function~\cite{argus} whose parameters are obtained by fitting the efficiencies determined at each $m$. A Gaussian function $G(\delta_{m},~\delta_{\sigma})$ is convolved to account for the detection resolution, where $\delta_{m}$ and $\delta_{\sigma}$ are the mean and the resolution of the Gaussian function, respectively. Its parameters $\delta_{m}$ and $\delta_{\sigma}$ are floated in the fit for $\chi_{cJ}$, but fixed for $\eta_c(2S)$ based on a control sample of $\chi_{cJ} \to p\bar{p}\pi^+\pi^-\pi^0$, yielding $\delta_m=-0.34\pm0.05$~MeV/$c^2$ and $\delta_{\sigma}=4.29\pm 0.14$~MeV/$c^2$.

The background lineshape of the $\psi(3686) \to p\bar{p}\pi^+\pi^-\pi^0(\gamma_\text{FSR})$ process is modeled using $f_\text{FSR}$-corrected MC simulated events, with its background yield left free in the fit. The lineshape for background events involving a misreconstructed radiative photon is taken from MC simulation, and its yield is constrained relative to the $\chi_{cJ}$ signal yields by a coefficient $f_{\rm mis\gamma}=0.014\pm 0.001$, defined as the ratio of the number of misreconstructed events to that of correctly reconstructed events in MC samples. The lineshape of the continuum background contribution is fixed, and its yield is controlled by the normalization factor $f_\text{norm}=8.53\pm0.12$. The uncertainties are predominantly statistical. All other residual background contributions are described by an Argus function~\cite{argus} with the background yield free. 
 
\begin{figure*}[htpb]
     \begin{center}
     \includegraphics[width=0.4\textwidth]{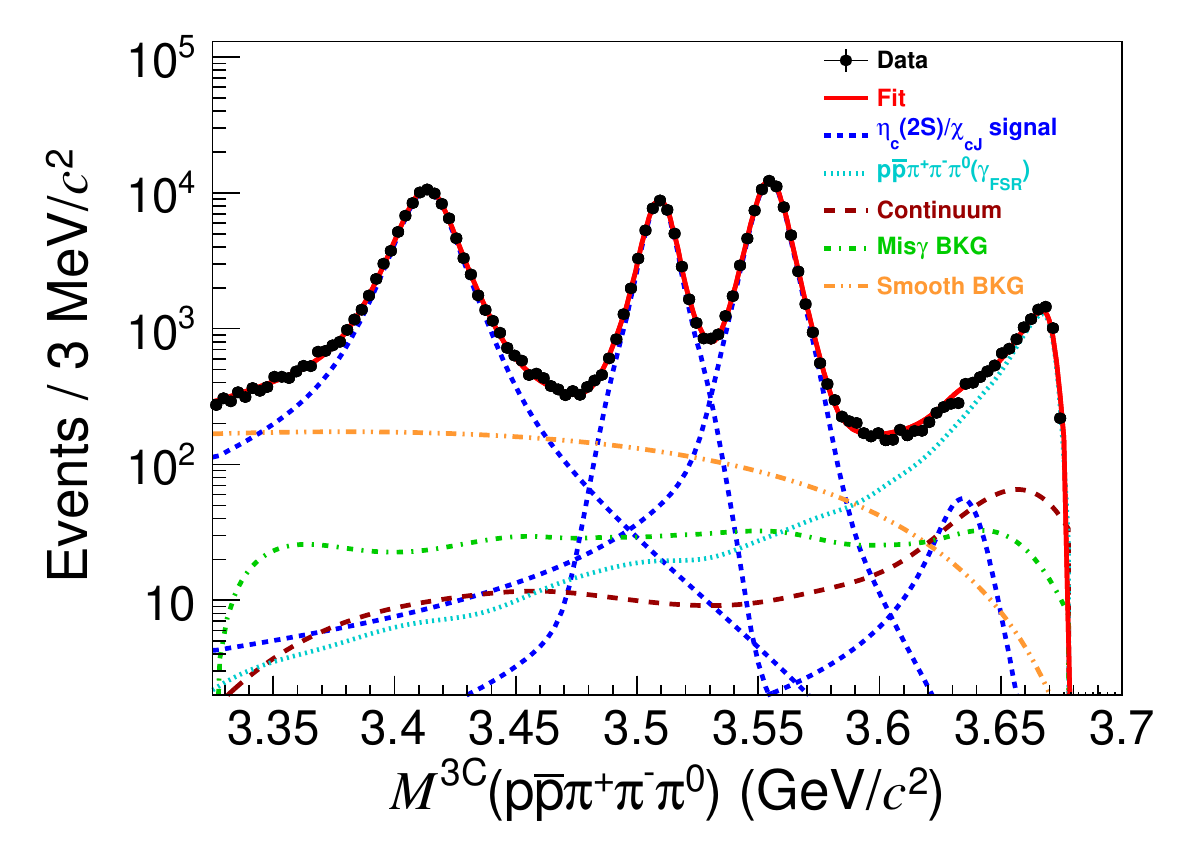}
     \includegraphics[width=0.4\textwidth]{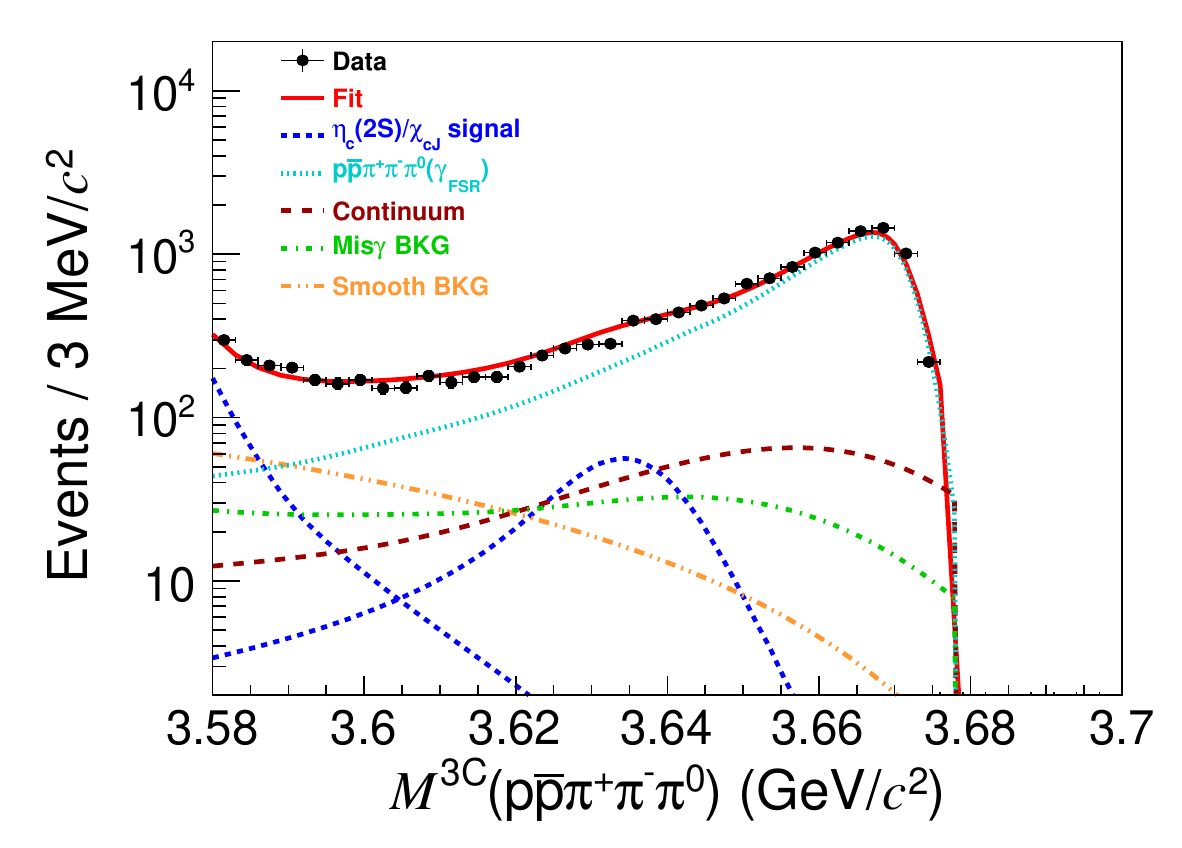}  
     \caption{Distributions of $M^\text{3C}(\pptripi)$ in data, presenting (left) the full fitting range and (right) the region containing the $\etacp$ signal. Data (dots with error bars) are presented together with the total fit result (red solid curve). The individual components include the $\eta_c$ and $\chi_{cJ}$ signals (blue dashed line), the $\psi(3686) \to p\bar{p}\pi^+\pi^-(\gamma_\text{FSR})$ process (cyan dotted line), the continuum background (brown long-dashed line), the background from misreconstructed radiative photons (green dash-dotted line), and a smooth background term (orange dash-dot-dotted line). The $\chi^2/\text{ndf}$ value of the fit is $169.8/106=1.6$, where ndf is the number of degrees of freedom.}
     \label{fig:secbf:subsec1:m3cpp3pi}
     \end{center}
 \end{figure*}

Figure~\ref{fig:secbf:subsec1:m3cpp3pi} shows the invariant mass distribution of $\pptripi$ after the 3C kinematic fit, $M^\text{3C}(\pptripi)$, from data together with the fit results, presenting both the full fitting range and the region containing the $\etacp$ signal. Table~\ref{tab:secbf:subsec1:fitresult} summarizes the extracted signal yields $N^{\rm extracted}_{\rm data}$ from the fit to the $M^\text{3C}(\pptripi)$ distribution and the detection efficiencies $\epsilon$ derived from signal MC samples. The $\chi^2/\text{ndf}$ value of the fit is $169.8/106=1.6$, where ndf is the number of degrees of freedom. The statistical significance for $\eta_c(2S) \to \pptripi$ is determined to be $4.1\sigma$, calculated from the difference in the logarithmic likelihoods $-2\ln(\mathcal{L}_{0}/\mathcal{L}_\text{max})$~\cite{significance}, where $\mathcal{L}_\text{max}$ and $\mathcal{L}_{0}$ represent the maximized likelihoods with and without the $\eta_{c}(2S)$ signal component, respectively, taking into account the difference in the number of degrees of freedom ($\Delta \text{ndf}$=1). The dominant systematic uncertainty, discussed in Sec.~\ref{sec:pptripi:subsys}, originates from the mass spectrum fitting procedure. Alternative fits to the $M^\text{3C}(p\bar{p}\pi^+\pi^-\pi^0)$ spectrum under varying conditions are performed; in all cases, the $\eta_c(2S)$ signal significance exceeds 3.3$\sigma$.

 \begin{table*}[htbp]
 \begin{center}
 \caption{Summary of the extracted signal yields, detection efficiencies, and branching fractions. The first uncertainties are statistical and the second systematic. The branching fraction marked with an asterisk ($*$) denotes the product branching fraction $\mathcal{B}[\psi(3686)\to \gamma \eta_{c}(2S)]\times\mathcal{B}[\eta_{c}(2S)\to p\bar{p}\pi^+\pi^-\pi^0]$.}
 \label{tab:secbf:subsec1:fitresult}
 \begin{tabular}{cccc}
 \hline \hline
 Channel & $N^{\rm extracted}_{\rm data}$ & $\epsilon$(\%) & Branching fraction \\ \hline 
        $\eta_{c}(2S) \to p\bar{p}\pi^{+}\pi^{-}\pi^{0}$ &$(5.43\pm0.58)\times10^2$ & 5.9 & $(3.4\pm 0.5\pm0.8)\times10^{-6~*}$ \\         
        $\chi_{c0} \to p\bar{p}\pi^{+}\pi^{-}\pi^{0}$ &$(9.86\pm 0.04)\times 10^4$ & 7.9 & $(4.79\pm 0.01\pm0.40)\times10^{-3}$  \\  
        $\chi_{c1} \to  p\bar{p}\pi^{+}\pi^{-}\pi^{0}$ & $(4.61\pm 0.03)\times 10^4$ & 8.3 & $(2.13\pm 0.01\pm0.17)\times10^{-3}$ \\  
        $\chi_{c2} \to  p\bar{p}\pi^{+}\pi^{-}\pi^{0}$ & $(7.07\pm 0.03)\times 10^4$  & 7.6 & $(3.72\pm 0.01\pm0.29)\times10^{-3}$ \\ \hline 
        $\chi_{c0} \to p\bar{p}\omega$ &$(1.05\pm0.02)\times10^4 $& 7.8 & $(5.76\pm0.01\pm0.42)\times 10^{-4}$ \\ 
        $\chi_{c1} \to p\bar{p}\omega$ &$(3.49\pm0.09)\times10^3 $& 8.1& $(1.85\pm0.01\pm0.13)\times 10^{-4}$ \\
        $\chi_{c2} \to p\bar{p}\omega$ &$(7.31\pm0.13)\times10^3 $& 7.2& $(4.51\pm0.01\pm0.33)\times 10^{-4}$ \\  \hline \hline
 \end{tabular}
 \end{center}
 \end{table*}

 The product of the branching fractions $\mathcal{B}[\psi(3686)\to \gamma \eta_{c}(2S)/\chi_{cJ}]\times\mathcal{B}[\eta_{c}(2S)/\chi_{cJ} \to p\bar{p}\pi^+\pi^-\pi^0]$ is calculated via
 \begin{linenomath*}
 \begin{equation}
   \frac{N^{\rm extracted}_{\rm data}}{N_{\psi(3686)}^{\rm tot}\cdot\epsilon \cdot \BF(\pi^0 \to \gamma\gamma)}, 
 \end{equation}
 \end{linenomath*}
 where $N_{\psi(3686)}^\text{tot}$ is the total number of $\psi(3686)$ events~\cite{psipNum} and $\BF(\pi^0 \to \gamma\gamma)$ is the branching fraction of the $\pi^0 \to \gamma\gamma$ decay. We employ the precisely measured branching fractions $\mathcal{B}[\psi(3686) \to \gamma \chi_{cJ}]$ to extract $\mathcal{B}(\chi_{cJ} \to p\bar{p}\pi^+\pi^-\pi^0)$ from the product branching fractions. The measured branching fractions are listed in Table~\ref{tab:secbf:subsec1:fitresult}. 

\subsection{\boldmath \texorpdfstring{$\chi_{cJ}\to p\bar{p}\omega$}{Lg}}
  We study the intermediate processes $\eta_c(2S)\to p\bar{p}\omega$ and $\chi_{cJ}\to p\bar{p}\omega$. Clear $\chi_{cJ}$ signals are observed, whereas no significant $\eta_c(2S)$ signal is seen in the invariant mass spectrum of $p\bar{p}\omega$ [$M(p\bar{p}\omega)$]. To investigate the contribution from $\chi_{cJ}\to p\bar{p}\omega$, a two-dimensional (2D) unbinned maximum likelihood fit is performed on the distributions of $M(p\bar{p}\pi^{+}\pi^{-}\pi^{0})(\equiv y)$ versus $M(\pi^{+}\pi^{-}\pi^{0})(\equiv x)$, where $M(\pi^{+}\pi^{-}\pi^{0})$ is the invariant mass of the $\pi^{+}\pi^{-}\pi^{0}$ system.
  
 \begin{figure*}[htbp]
     \begin{center}
     \includegraphics[width=0.4\textwidth]{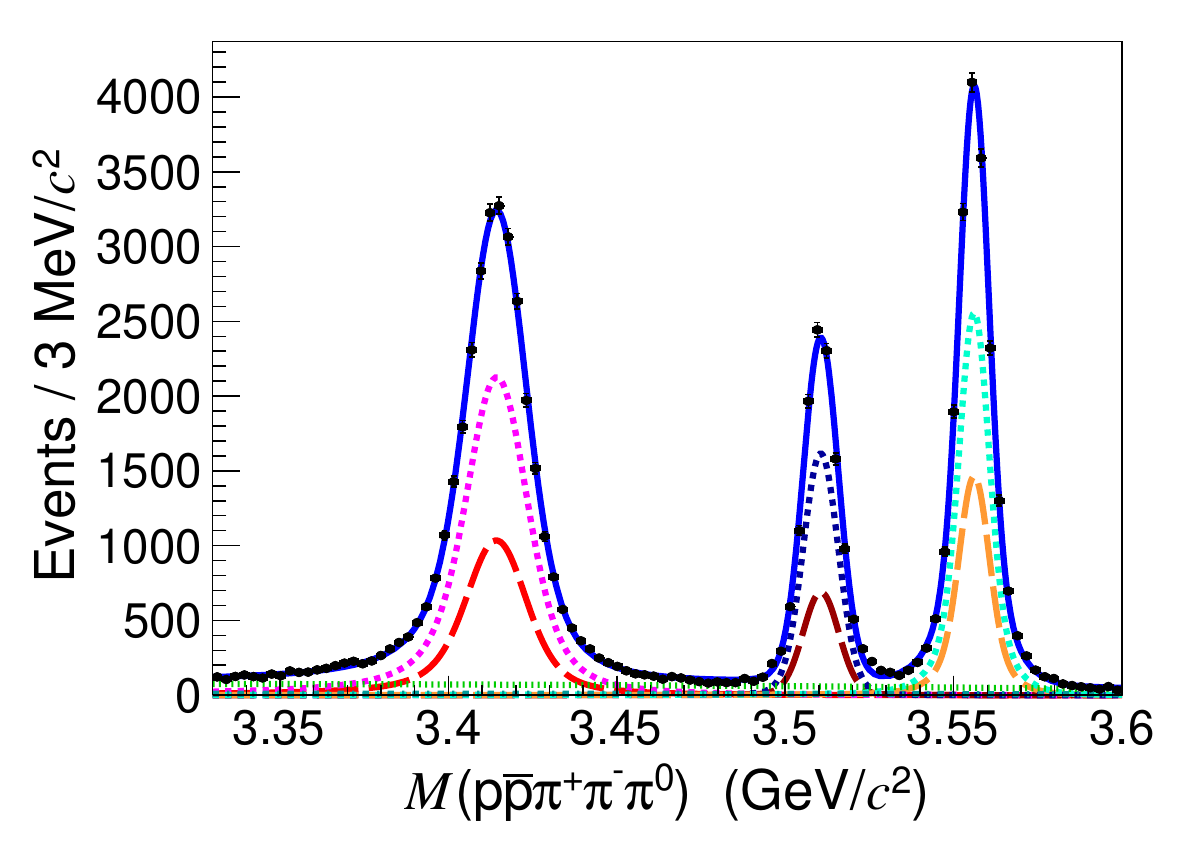}
     \includegraphics[width=0.4\textwidth]{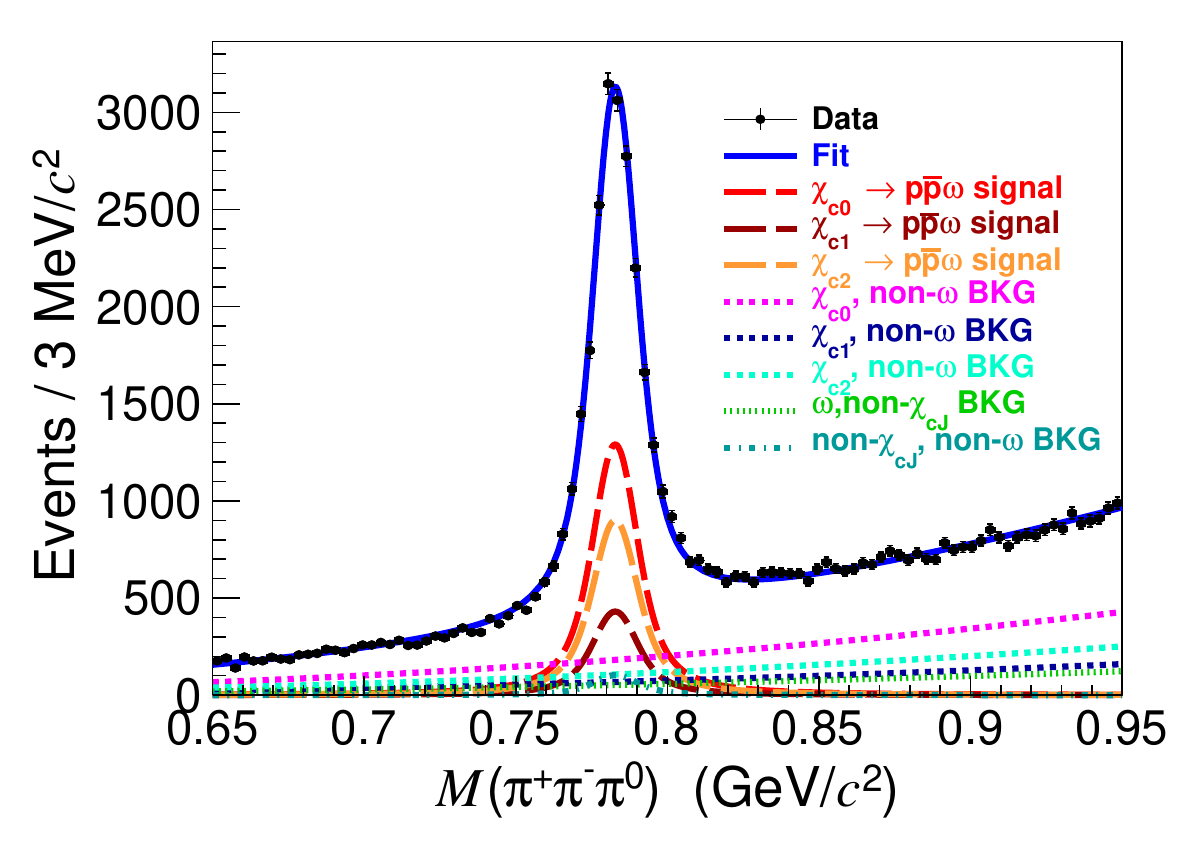}  
     \caption{Projections of the 2D distributions of (left) $M(p\bar{p}\pi^{+}\pi^{-}\pi^{0})$ versus (right) $M(\pi^{+}\pi^{-}\pi^{0})$ in data. Data (dots with error bars) are presented together with the total fit result (blue solid curve). The individual components include the $\chi_{cJ} \to p\bar{p}\omega$ signals (red, brown, and orange long-dashed lines for $J=0,~1,~2$, respectively), the $\chi_{cJ}$ but non-$\omega$ background (pink, dark-blue, and teal dashed lines for $J=0,~1,~2$, respectively), the $\omega$ but non-$\chi_{cJ}$ background (green dotted line), and the non-$\omega$ and non-$\chi_{cJ}$ background (cyan dash-dotted line).} 
     \label{fig:secbf:subsec2:mpp3pi}
     \end{center}
 \end{figure*}

In the 2D fit, the probability density function (PDF) of the signal is constructed as $\mathcal{S}_x \otimes \mathcal{S}_y$, where $\mathcal{S}_x$ and $\mathcal{S}_y$ describe the $\omega$ signal along the $x$ dimension and the $\chi_{cJ}$ signal along the $y$ dimension, respectively. The $\mathcal{S}_x$ is a BW function convolved with a Gaussian resolution function, with all parameters left free in the fit. The form of $\mathcal{S}_y$ follows Eq.~(\ref{sigsp}), in which the parameters of the Gaussian resolution are determined from a 1D fit to the $M(p\bar{p}\pi^{+}\pi^{-}\pi^{0})$ distribution and subsequently held fixed in the 2D fit. The background events are categorized into three components. The first category consists of events originating from decays that are associated with neither the $\omega$ nor the $\chi_{cJ}$. Its PDF is constructed as $\mathcal{P}_x \otimes \mathcal{P}_y$, where $\mathcal{P}_x$ and $\mathcal{P}_y$ are second-order polynomial functions characterizing the combinatorial background shapes in the $M(\pi^{+}\pi^{-}\pi^{0})$ and $M(p\bar{p}\pi^{+}\pi^{-}\pi^{0})$ distributions, respectively. The PDFs of the second and third background components are $\mathcal{S}_x \otimes \mathcal{P}_y$ and $\mathcal{S}_y \otimes \mathcal{P}_x$, describing events from the $\omega$ decay but not from the $\chi_{cJ}$ decay, and events from the $\chi_{cJ}$ decay but not from the $\omega$ decay, respectively.

 Figure~\ref{fig:secbf:subsec2:mpp3pi} shows the one-dimensional (1D) projections of the 2D fit. The $\chi^2/\text{ndf}$ values are $135.8/86=1.6$ and $102.3/86=1.2$ for the fits to the $M(p\bar{p}\pi^+\pi^-\pi^0)$ and $M(\pi^+\pi^-\pi^0)$ distributions of Fig.~\ref{fig:secbf:subsec2:mpp3pi}, respectivaly. The branching fractions of $\chi_{cJ} \to p\bar{p}\omega$ are calculated via
 \begin{linenomath*}
 \begin{equation}
   \mathcal{B}(\chi_{cJ}\to p\bar{p}\omega)=\frac{N^{\rm extracted}_{\rm data}}{N_{\psi(3686)}^{\rm tot}\cdot\epsilon\cdot \mathcal{B}^{\prime}},
 \end{equation}
 \end{linenomath*}
where $\mathcal{B}^{\prime}=\mathcal{B}[\psi(3686)\to \gamma \chi_{cJ}]\cdot \mathcal{B}(\omega \to \pi^+\pi^-\pi^0)\cdot \mathcal{B}(\pi^0 \to \gamma \gamma)$. Table~\ref{tab:secbf:subsec1:fitresult} lists the extracted signal yields of $\chi_{cJ} \to p\bar{p}\omega$, the detection efficiencies, and the measured branching fractions. 
     
%%%%%%%%%%%%%%%%%%%%%%%%%%%%%%%%%%%%%%%%%%%%%%%%%%%%%%%%%%%%%%%%%%%%%
\section{\label{sec:syserr} Systematic Uncertainties}
\label{sec:pptripi:subsys}

Table~\ref{tab:syserr} summarizes the sources of systematic uncertainty in the branching fraction measurements. The total systematic uncertainty is obtained by adding all individual sources in quadrature. Each source is described below.

 \begin{table*}[htpb]
 \begin{center}
 \caption{\label{tab:syserr} Systematic uncertainties (in unit of \%) in the measurements of the branching fractions. The $f$ and $f^{\prime}$ represent $\pptripi$ and $p\bar{p}\omega$, respectively. The symbol ``--'' denotes a negligible uncertainty. }
 \begin{tabular}{c c c c c c c c}
 \hline\hline
    Source & $\etacp \to f$ & $\chi_{c0} \to f$ & $\chi_{c1} \to f$ & $\chi_{c2} \to f$ & $\chi_{c0} \to f^{\prime}$ & $\chi_{c1} \to f^{\prime}$ & $\chi_{c2} \to f^{\prime}$ \\ \hline
    Tracking & 4.0 & 4.0  & 4.0  &  4.0 & 4.0 & 4.0  &  4.0   \\ 
    PID         & 4.0  & 4.0  & 4.0   & 4.0 & 4.0   & 4.0   & 4.0  \\
    Photon reconstruction  & 3.0 & 3.0 & 3.0   &3.0  & 3.0  & 3.0  & 3.0 \\
    Helix parameter correction  & 0.9 & 0.8  & 0.9  & 0.8  & 0.7  & 0.9 & 0.8 \\
    Background veto & 10.5 & 0.6 & 1.3 & 0.6 & 0.5 & 1.2 & 0.6 \\
    Mass of $\eta_{c}(2S)$ &2.3 & 0.1 & 0.1 & 0.1 & - & - & - \\
    Width of $\eta_{c}(2S)$ & 4.3 & 0.1 & 0.1 & 0.1 & - & - & - \\
    Efficiency curve & 4.6 & 0.4 & 0.3 & 0.6 & 0.1 & 0.1 & 0.1 \\
    Damping function & 10.2 & 1.0 & 0.8 & 0.6 & 1.0 & 1.0 & 1.0  \\
    Gaussian resolutions & 0.1 & 0.1 & 0.1 & 0.1 & 0.1 & 0.1 & 0.1  \\
    Normalization factor   & 2.6 & 0.1 & 0.1 & 0.1 & - & - & - \\
    FSR factor &14.0 & 0.1 & 0.1 & 0.1 & - & - & - \\
    Mis-reconstructed coefficient  & 1.8 & 4.5 & 3.4 & 3.1 & - & - & - \\
    Background lineshape  & 0.3 & 0.1 & 0.2 & 0.1 &0.2 &0.2 &0.1 \\
    Fitting range  &1.3 & 0.6  & 0.3  &0.1 &1.2 &0.6  &1.9 \\
    Data-MC discrepancy & 3.8 & - & - & - & - & - & - \\
    Size of MC sample  & 0.2 & 0.2 & 0.2 & 0.2 & 0.2 & 0.2 & 0.2 \\
    Quoted branching fraction & - & 2.4 & 2.8  &2.5 & 2.5 & 2.9 &2.6 \\ 
    Number of $\psi(3686)$ events & 0.5 & 0.5 & 0.5 & 0.5 & 0.5 & 0.5 & 0.5 \\ \hline
    Total & 22.9 & 8.4 & 8.0 & 7.7 & 7.2 & 7.3 & 7.4 \\ \hline\hline
 \end{tabular}
 \end{center}
 \end{table*}

The systematic uncertainty associated with the tracking efficiency is evaluated with a control sample of $J/\psi \to p\bar{p}\pi^+\pi^-$, and a $1.0\%$ uncertainty is assigned per pion or proton~\cite{tracking-error1, tracking-error2}. The systematic uncertainty arising from proton PID is studied using the control sample $e^+e^-\to p\bar{p}\pi^0$ and is estimated to be $1.0\%$ per proton~\cite{proton-pid}. For pion PID, a $1.0\%$ uncertainty per pion is determined based on the control sample $J/\psi \to p\bar{p}\pi^+\pi^-$~\cite{tracking-error1}. The uncertainty due to photon reconstruction is taken as $1.0\%$ per photon, derived from studies of photon detection efficiency using the control samples $J/\psi \to \pi^+\pi^-\pi^0$ and $e^+ e^- \to \gamma \gamma$~\cite{photonRec1,photonRec2}.

A helix-parameter correction has been applied to the MC-simulated samples for each track. The correction factors for protons and pions are studied using the control samples $e^+ e^- \to p\bar{p}\pi^+\pi^-$ and $\psi(3686) \to \gamma \chi_{c0},~ \chi_{c0} \to 3(\pi^+\pi^-)$~\cite{helixfit1,3cfit}. The systematic uncertainty due to the helix-parameter correction is taken as half the difference between the detection efficiencies with and without the track helix-parameter correction.

The systematic uncertainties associated with background vetoes encompass those from cross-feed suppression and the rejection of specific background components, namely $\psi(3686) \to \pi^+\pi^-J/\psi$, $\psi(3686) \to \pi^0\pi^0J/\psi$, and $\Lambda/\Sigma$-related processes. The uncertainty due to cross-feed suppression is evaluated by varying the mass window requirements for $\left|M(\gamma_{\pi^0,1}\gamma_\text{rad}) - m(\pi^0)\right|$ from 13 to 17~MeV/$c^2$ and for $\left|M(\gamma_{\pi^0,2}\gamma_\text{rad}) - m(\pi^0)\right|$ from 11 to 14~MeV/$c^2$, in steps of 0.5~MeV/$c^2$, following the method of the Barlow test in Ref.~\cite{barlow}. The resulting systematic uncertainty is taken as the maximum difference in the branching fractions. For the veto on the $\psi(3686) \to \pi^+\pi^-J/\psi$ background events, the requirement on $\left|RM(\pi^{+}\pi^{-}) - m(J/\psi)\right|$ is varied from $>$5.4~MeV/$c^2$ to $>$6.8~MeV/$c^2$ in intervals of 0.2~MeV/$c^2$. The maximum deviation from the nominal branching fraction is assigned as the corresponding systematic uncertainty. Similarly, the uncertainty from vetoing the $\psi(3686) \to \pi^0\pi^0J/\psi$ events is estimated by varying the requirement on $\left|M(p\bar{p}\pi^+\pi^-) - m(J/\psi)\right|$ from $>$21.5~MeV/$c^2$ to $>$24.5~MeV/$c^2$ in steps of 0.5~MeV/$c^2$, with the maximum deviation taken as the systematic uncertainty. The uncertainty related to suppressing the $\Lambda/\Sigma$-related background events is assessed by varying the mass-window requirements for $M(p\pi^-)$ from 16 to 20~MeV/$c^2$ with an interval of 1~MeV/$c^2$, for $M(p\pi^0)$ from 24 to 32~MeV/$c^2$ with an interval of 2~MeV/$c^2$, and for $M(\gamma\Lambda)$ from 20 to 28~MeV/$c^2$ with an interval of 2~MeV/$c^2$. The largest difference in the branching fractions relative to the nominal one for each background veto is assigned as the systematic uncertainty. The total systematic uncertainty from all background veto sources is obtained by adding the individual contributions in quadrature.

The systematic uncertainties due to the fixed mass and width of $\eta_{c}(2S)$ are evaluated by varying their known values within their respective $\pm1\sigma$ uncertainties~\cite{pdg}. The maximum differences in the extracted signal yields are assigned as the corresponding systematic uncertainties. 

The systematic uncertainty associated with the efficiency curve is evaluated by varying its parameters by $\pm 1$ standard deviation and assigning the maximum difference in the extracted signal yields as the systematic uncertainty. To assess the uncertainty from the choice of damping function, the alternative form $f_d(E_{\gamma})=$ exp$(-E_{\gamma}^2/8\beta^2)$ employed by the CLEO Collaboration~\cite{cleo} is adopted, with the parameter $\beta$ fixed at $65.0$~MeV. The resulting difference in the extracted signal yields relative to the nominal value is taken as the systematic uncertainty. The uncertainty arising from the Gaussian resolution is estimated by varying its parameters within their $\pm1$ standard deviation statistical errors, with the maximum difference in the extracted signal yields assigned as the systematic uncertainty.

The systematic uncertainties related to background contributions in the measurements of $\mathcal{B}[\eta_c(2S)/\chi_{cJ} \to \pptripi]$ arise from the normalization factor $f_\text{norm}$, the FSR factor $f_\text{FSR}$, and the misreconstruction coefficient $f_{\rm mis\gamma}$. Each factor is varied within its $\pm1\sigma$ statistical uncertainty, and the maximum difference in the extracted signal yields is assigned as the corresponding systematic uncertainty. For the measurements of $\mathcal{B}(\chi_{cJ} \to p\bar{p}\omega)$, the uncertainty due to the background lineshape is evaluated by replacing the nominal second-order polynomial with a third-order polynomial. The difference in the extracted signal yields is taken as the systematic uncertainty.

The systematic uncertainties associated with the background lineshape of the FSR process or with the misreconstructed radiative photon originate from the finite MC statistics. Doubling the background statistics and refitting the data leads to negligible changes in the signal yields. The impact of the direct decay $\psi(3686) \to p\bar{p}\pi^+\pi^-\pi^0(\gamma_\text{FSR})$ is estimated with intermediate resonances taken into account, and the corresponding systematic uncertainty is at most 0.3\%.

To assess the systematic uncertainty arising from the choice of fitting range in the measurements of $\mathcal{B}[\eta_c(2S)/\chi_{cJ} \to \pptripi]$, the lower bound of the $M(p\bar{p}\pi^+\pi^-\pi^{0})$ interval is varied from 3.320 to 3.328~GeV/$c^2$ in steps of 2~MeV/$c^2$. The maximum observed variation in the extracted signal yields is assigned as the systematic uncertainty. For the measurements of $\mathcal{B}(\chi_{cJ} \to p\bar{p}\omega)$, the uncertainty due to the 2D fitting range is evaluated by independently varying the bounds of both invariant mass windows. The nominal $\pi^+\pi^-\pi^0$ mass window $(0.65, 0.95)$~GeV/$c^2$ is shifted by $\pm0.015$~GeV/$c^2$ to $(0.635, 0.965)$~GeV/$c^2$. Simultaneously, the nominal $p\bar{p}\pi^+\pi^-\pi^0$ mass window $(3.33, 3.60)$~GeV/$c^2$ is adjusted by $\pm0.015$~GeV/$c^2$, resulting in the window $(3.345, 3.585)$~GeV/$c^2$. The largest deviation from the nominal branching fraction is taken as the systematic uncertainty.   

In the $\chi_{cJ} \to \pptripi$ processes, slight discrepancies are observed between data and MC simulation in the momentum distributions of $\pi^+$, $\pi^-$, and $\pi^0$. These discrepancies are attributed to the absence of intermediate resonances such as $\omega/\eta \to \pi^+\pi^-\pi^0$ in the signal MC sample. Furthermore, clear intermediate states, $\omega$ and $\eta$, are observed in the $M(\pi^+\pi^-\pi^0)$ distribution in data, but are absent in the MC simulation. 
To account for these discrepancies, we correct the detection efficiency using a reweighting procedure based on the three-dimensional momentum distributions of $\pi^+$, $\pi^-$, and $\pi^0$, and the 1D distribution of $M(\pi^+\pi^-\pi^0)$ in data. In addition, the detection efficiency of $\chi_{cJ} \to p\bar{p}\omega$ is corrected using the three-dimensional momentum distributions of $p$, $\bar{p}$, and $\omega$. 

For the $\eta_c(2S) \to \pptripi$ decay, limited signal statistics preclude a direct data-driven efficiency correction. Instead, we conservatively estimate the associated systematic uncertainty using $\chi_{cJ} \to \pptripi$ as a control sample. The difference in detection efficiencies with and without correction is calculated for the $\chi_{cJ}$ samples. This difference is then scaled to $\eta_c(2S)$ by assuming a linear relationship with respect to $\chi_{cJ}$ and $\eta_c(2S)$ and performing a linear fit. The relative difference for $\eta_c(2S)$ is then calculated from the linear fit as a conservative estimate of the systematic uncertainty for $\eta_c(2S)$.

The systematic uncertainty due to the finite statistics of the MC sample is evaluated via $\sqrt{(1-\epsilon)/\epsilon N}$, where $N$ is the number of generated signal MC events. The uncertainties associated with the quoted branching fractions for $\psi(2S) \to \gamma\chi_{cJ}$, $\omega \to \pi^+\pi^-\pi^0$, and $\pi^0 \to \gamma\gamma$ are taken from their respective world-average values~\cite{pdg}. The uncertainty on the total number of $\psi(3686)$ events is taken to be 0.5\%, as documented in Ref.~\cite{psipNum}.  

%%%%%%%%%%%%%%%%%%%%%%%%%%%%%%%%%%%%%%%%%%%%%%%%%%%%%%%%%%%%%%%%%%%%%
\section{Summary} 

Based on a sample of $(2.712\pm0.014)\times 10^{9}$ $\psi(3686)$ events collected with the BESIII detector at the BEPCII collider, we investigate the radiative decay $\psi(3686)\to\gamma p\bar{p}\pi^{+}\pi^{-}\pi^{0}$. Evidence for the decay $\eta_{c}(2S) \to p\bar{p}\pi^+\pi^-\pi^0$ is reported for the first time with a signal significance of 3.3$\sigma$. The decays $\chi_{cJ} \to p\bar{p}\pi^+\pi^-\pi^{0}$ are also observed. The first measured branching fractions are $\mathcal{B}[\psi(3686)\to \gamma \eta_{c}(2S)]\times\mathcal{B}[\eta_{c}(2S) \to p\bar{p}\pi^+\pi^-\pi^0] = (3.4\pm 0.5\pm0.8) \times 10^{-6}$, $\mathcal{B}(\chi_{c0} \to \pptripi) = (4.79\pm 0.01\pm0.40) \times 10^{-3}$, $\mathcal{B}(\chi_{c1} \to \pptripi) = (2.13\pm 0.01\pm0.17) \times 10^{-3}$, and $\mathcal{B}(\chi_{c2} \to \pptripi) = (3.72\pm 0.01\pm0.29) \times 10^{-3}$. 

In addition, we report improved measurements of the branching fractions of $\chi_{cJ}\to p\bar{p}\omega$ decays: $\mathcal{B}(\chi_{c0} \to p\bar{p}\omega) = (5.76\pm0.01\pm0.42)\times10^{-4}$, $\mathcal{B}(\chi_{c1} \to p\bar{p}\omega) = (1.85\pm0.01\pm0.13)\times10^{-4}$, and $\mathcal{B}(\chi_{c2} \to p\bar{p}\omega) = (4.51\pm0.01\pm0.33)\times10^{-4}$. The measured branching fractions for $\chi_{c0}$ and $\chi_{c1}$ are consistent with the world-averaged values~\cite{pdg} within one standard deviation. For $\chi_{c2}$, the measured value agrees with the world-averaged value~\cite{pdg} within two standard deviations. The measurements show a 30\%--50\% improvement in precision relative to the world-averaged values.

These measurements not only enrich the experimental knowledge of $\eta_c(2S)$ and $\chi_{cJ}$ hadronic decays but also contribute to a more coherent understanding of the charmonium spectrum below the open-charm threshold. Additionally, this measurement of $\eta_c(2S)$ decays could provide valuable experimental information to test the analogous ``12\% rule'' for $\eta_c(1S)$ and $\eta_c(2S)$. \\

%%%%%%%%%%%%%%%%%%%%%%%%%%%%%%%%%%%%%%%%%%%%%%%%%%%%%%%%%%%%%%%%%%%%%
\acknowledgments

The BESIII Collaboration thanks the staff of BEPCII (https://cstr.cn/31109.02.BEPC) and the IHEP computing center for their strong support. This work is supported in part by National Key R\&D Program of China under Contracts Nos. 2025YFA1613900, 2023YFA1606000, 2023YFA1606704, 2025YFA1613900; National Natural Science Foundation of China (NSFC) under Contracts Nos. 11635010, 11935015, 11935016, 11935018, 12025502, 12035009, 12035013, 12061131003, 12192260, 12192261, 12192262, 12192263, 12192264, 12192265, 12221005, 12225509, 12235017, 12342502, 12361141819, 12535005; the Chinese Academy of Sciences (CAS) Large-Scale Scientific Facility Program; the Strategic Priority Research Program of Chinese Academy of Sciences under Contract No. XDA0480600; CAS under Contract No. YSBR-101; 100 Talents Program of CAS; The Institute of Nuclear and Particle Physics (INPAC) and Shanghai Key Laboratory for Particle Physics and Cosmology; ERC under Contract No. 758462; German Research Foundation DFG under Contract No. FOR5327; Istituto Nazionale di Fisica Nucleare, Italy; Knut and Alice Wallenberg Foundation under Contracts Nos. 2021.0174, 2021.0299, 2023.0315; Ministry of Development of Turkey under Contract No. DPT2006K-120470; National Research Foundation of Korea under Contract No. NRF-2022R1A2C1092335; National Science and Technology fund of Mongolia; Polish National Science Centre under Contract No. 2024/53/B/ST2/00975; STFC (United Kingdom); Swedish Research Council under Contract No. 2019.04595; U. S. Department of Energy under Contract No. DE-FG02-05ER41374.

%%%%%%%%%%%%%%%%%%%%%%%%%%%%%%%%%%%%%%%%%%%%%%%%%%%%%%%

%

\end{document}